\documentclass[11pt,letterpaper]{article}
\usepackage{emnlp2017}
\usepackage{times}
\usepackage{latexsym}
\usepackage{amsmath}
\usepackage{graphicx}
\usepackage[]{algorithm2e}
\usepackage{url}

\usepackage[font=small]{caption}

\emnlpfinalcopy

\def\emnlppaperid{1419}

\title{Temporal Information Extraction for Question Answering\\ 
Using Syntactic Dependencies in an LSTM-based Architecture}

\author{Yuanliang Meng, Anna Rumshisky, Alexey Romanov \\
        {\tt\small \{ymeng,arum,aromanov\}@cs.uml.edu} \\
        Department of Computer Science \\
        University of Massachusetts Lowell \\
        Lowell, MA 01854
}

\date{}

\begin{document}
\maketitle
\begin{abstract}

  In this paper, we propose to use a set of simple, uniform in architecture LSTM-based models to recover different kinds of temporal relations from text. Using the shortest dependency path between entities as input, the same architecture is implemented to extract intra-sentence, cross-sentence, and document creation time relations. A ``double-checking'' technique reverses entity pairs in classification, boosting the recall of positive cases and reducing misclassifications between opposite classes. An efficient pruning algorithm resolves conflicts globally. Evaluated on QA-TempEval (SemEval2015 Task 5), our proposed technique outperforms state-of-the-art methods by a large margin. We also conduct intrinsic evaluation and post state-of-the-art results on Timebank-Dense.
  



\end{abstract}

\section{Introduction}

Recovering temporal information from text is essential to many text processing tasks that require deep language understanding, such as answering questions about the timeline of events or automatically producing text summaries. This work presents intermediate results of an effort to build a temporal reasoning framework with contemporary deep learning techniques.

Until recently, there has been remarkably few attempts to evaluate temporal information extraction (TemporalIE) methods in context of downstream applications that require reasoning over the temporal representation.  One recent effort to conduct such evaluation was SemEval2015 Task 5, a.k.a. QA-TempEval \cite{Llorens2015:TempEval}, which used question answering (QA) as the target application.  QA-TempEval evaluated systems producing TimeML \cite{Pustejovsky03timeml:robust} annotation based on how well their output could be used in QA.  We believe that 
application-based evaluation of TemporalIE should eventually completely replace the intrinsic evaluation if we are to make progress, and therefore we evaluated our techniques mainly using QA-TempEval setup.  

Despite the recent advances produced by multi-layer neural network architectures in a variety of areas, the research community is still struggling to make neural architectures work for linguistic tasks that require long-distance dependencies (such as discourse parsing or coreference resolution).  Our goal was to see if a relatively simple architecture with minimal capacity for retaining information was able to incorporate the information required to identify temporal relations in text.  

Specifically, we use several simple LSTM-based components to recover ordering relations between temporally relevant entities (events and temporal expressions).  These components are fairly uniform in their architecture, relying on dependency relations recovered with a very small number of mature, widely available processing tools, and require minimal engineering otherwise. 
To our knowledge, this is the first attempt to apply such simplified techniques to the TemporalIE task, and we  demonstrate this streamlined architecture is able to outperform state-of-the-art results on a temporal QA task with a large margin.  

In order to demonstrate generalizability of our proposed architecture, we also evaluate it intrinsically using TimeBank-Dense\footnote{\url{https://www.usna.edu/Users/cs/nchamber/caevo/##corpus}} \cite{chambers2014dense}.  TimeBank-Dense annotation aims to approximate a complete temporal relation graph by including all intra-sentential relations, all relations between adjacent sentences, and all relations with document creation time.  Although our system was not optimized for such a paradigm, and this data is quite different in terms of both the annotation scheme and the evaluation method, we obtain state-of-the-art results on this corpus as well.

\section{Related Work}

A multitude of TemporalIE systems have been developed over the past decade both in response to the series of shared tasks organized by the community \cite{verhagen2007semeval,verhagen2010semeval,uzzaman2012tempeval,sun2013evaluating,bethard2015semeval,llorenssemeval,minardsemeval} and in standalone efforts \cite{chambers2014dense,DBLP:journals/corr/Mirza16}.
%
%

The best methods used by TemporalIE systems to date tend to rely on highly engineered task-specific models using traditional statistical learning, typically used in succession \cite{sun2013evaluating,chambers2014dense}.  For example, in a recent QA-TempEval shared task, the participants routinely used a series of classifiers (such as support vector machine (SVM) or  hidden Markov chain SVM) or hybrid methods combining hand crafted rules and SVM, as was used by the top system in that challenge \cite{hlt-fbk}.
While our method also relies on decomposing the temporal relation extraction task into subtasks, we use essentially the same simple LSTM-based architecture for different components, that consume a highly simplified representation of the input.


Although there has not been much work applying deep learning techniques to TemporalIE,  some relevant work has been done on a similar (but typically more local) task of relation extraction.  Convolutional neural networks \cite{Zeng:2014} and recurrent neural networks both have been used for argument relation classification and similar tasks \cite{DBLP:journals/corr/ZhangW15a,Xu2015,DBLP:journals/corr/VuAGS16}.  We take inspiration from some of this work, including specifically the approach proposed by \newcite{Xu2015} which uses syntactic dependencies.





\section{Dataset}

We used QA-TempEval (SemEval 2015 Task 5)\footnote{\url{http://alt.qcri.org/semeval2015/task5/}} data and evaluation methods in our experiments. The training set contains 276 annotated TimeML files, mostly news articles from major agencies or Wikinews 
from late 1990s to early 2000s. 
This data contains annotations for events, temporal expressions (referred to as \textsc{timex}es), and temporal relations (referred to as \textsc{tlink}s).
The test set contains unannotated files in three genres: 10 news articles composed in 2014, 10 Wikipedia articles about world history, and 8 blogs entries from early 2000s. 


In QA-TempEval, evaluation is done via a QA toolkit which contains yes/no questions about temporal relations between two events or an event and a temporal expression. 
QA evaluation is not available for most of the training data except for 25 files, for which 79 questions are available.  We used used this subset of the training data for validation.
%
The test set contains unannotated files, so QA is the only way to measure the performance. The total of 294 questions is available for the test data, see Table~\ref{test-data}. 
%
%
%

We also use TimeBank-Dense dataset, which contains  
a subset of the documents in QA-TempEval. In TimeBank-Dense, all entity pairs in the same sentence or in consecutive sentences are labeled. If there is no information about the relation between two entities, it is labeled as ``vague''. We follow the experimental setup in \cite{chambers2014dense}, which splits the corpus into training/validation/test sets of 22, 5, and 9 documents, respectively.

\begin{figure*}[ht]

\begin{center}
\includegraphics[width=0.8\linewidth, clip, trim=0.5cm 0cm 0.5cm 0.8cm]{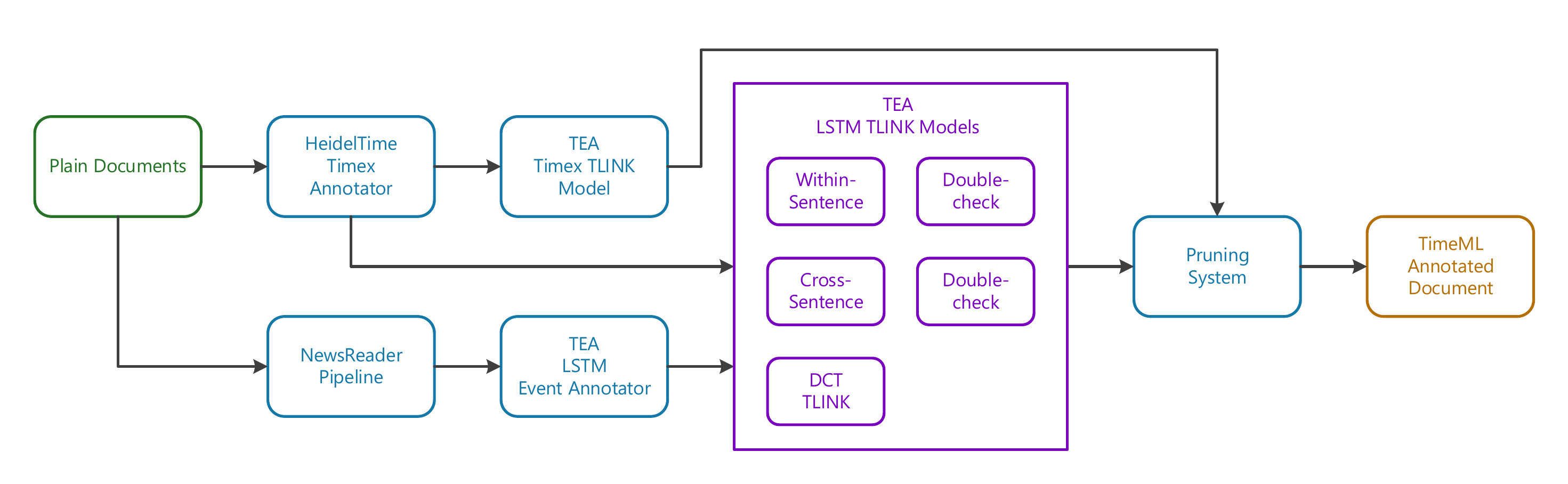}
\end{center}
\vspace{-1.5em}
\caption{System overview for our temporal extraction annotator (TEA) system}

\label{fig:system}

\end{figure*}


\section{\textsc{Timex} and Event Extraction}

The first task in our TemporalIE pipeline (TEA) is to identify time expressions (\textsc{timex}es) and events in text. We utilized the HeidelTime package \cite{StroetgenGertz2013:LREjournal} to identify \textsc{timex}es.
%
%
%
%
%
We trained a neural network model to identify event mentions.  
Contrary to common practice in TemporalIE, our models do not rely on event attributes, and thus we did not attempt to identify them.


\begin{table}[ht]
\begin{center}
\resizebox{\hsize}{!}{
\begin{tabular}{|l|l|}
\hline \bf Feature & \bf Explanation \\ \hline
is\_main\_verb & whether the token is the main verb of a sentence \\
is\_predicate & whether the token is the predicate of a phrase \\
is\_verb & whether the token is a verb \\
is\_noun & whether the token is a noun \\
\hline
\end{tabular}
}
\end{center}
\vspace{-1em}
\caption{\label{token-features} Token features for event extraction }
\end{table}

We perform tokenization, part-of-speech tagging, and dependency parsing using  NewsReader \cite{Agerri2014:newsreader}. Every token is represented with a set of features derived from preprocessing. 
Syntactic dependencies are not used for event extraction, but are used later in the pipeline for \textsc{tlink} classification.
The features used to identify events are listed in Table \ref{token-features}. 



\begin{figure}[ht]
\begin{center}
\includegraphics[width=1.0\linewidth, clip, trim=1.5cm 1.3cm 2.3cm 1cm]{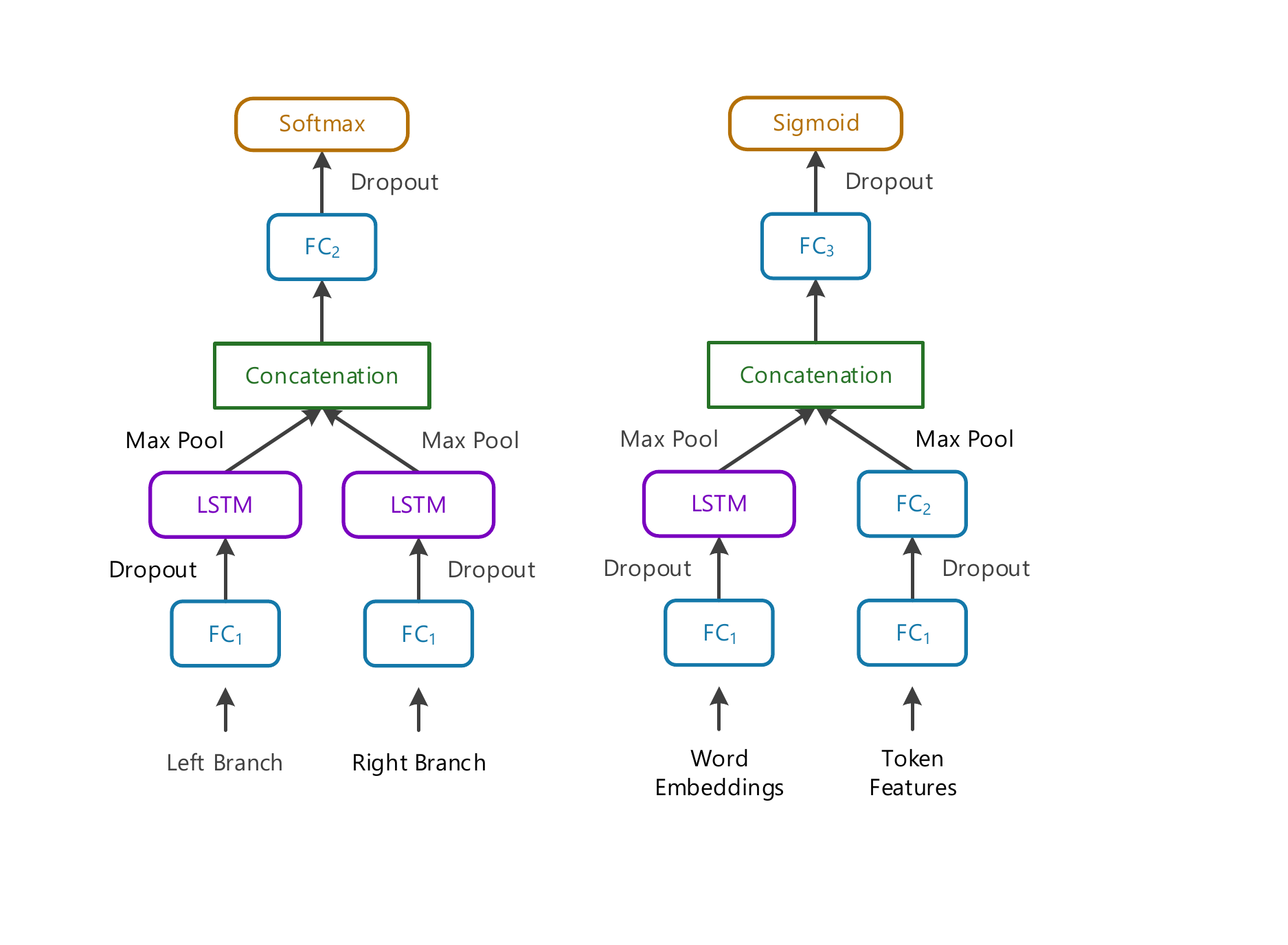}
\end{center}
\vspace{-1.5em}
\caption{Model architecture. Left: intra-sentence and cross-sentence model. Right: Event extraction model.}
\label{fig:rnn_model}
\end{figure}

The event extraction model uses long short-term memory (LSTM) \cite{Hochreiter:LSTM}, an RNN architecture well-suited for sequential data. 
The extraction model has two components, as shown on the right of Figure \ref{fig:rnn_model}. One component is an LSTM layer which takes word embeddings as input. The other component takes 4 token-level features as input. These components produce hidden representations which are concatenated, and fed into an output layer which performs binary classification. 
For each token, we use four tokens on each side to represent the surrounding context. The resulting sequence of nine word embeddings is then used as input to an LSTM layer. If a word is near the edge of a sentence, zero padding is applied.
We only use the token-level features of the target token, and ignore those from the context words. The 4 features are all binary, as shown in Table \ref{token-features}.
Since the vast majority of event mentions in the training data are single words, we only mark single words as event mentions. 



\section{\textsc{Tlink} Classification}
\label{sec:tlink-classification}

Our temporal relation (\textsc{tlink}) classifier consists of four components: an LSTM-based model for intra-sentence entity relations, an LSTM-based model for cross-sentence relations, another LSTM-based model for relations with document creation time, and a rule-based component for \textsc{timex} pairs. The four models perform \textsc{tlink} classifications independently, and the combined results are fed into a pruning module to remove the conflicting \textsc{tlink}s. 
The three LSTM-based components use the same streamlined architecture over token sequences recovered from shortest dependency paths between entity pairs.

\subsection{Intra-Sentence Model}
\label{sect:within}



A \textsc{tlink} extraction model should be able to learn the patterns that correspond to specific temporal relations, such as specific temporal prepositional phrases and clauses with temporal conjunctions.  This suggests such models may benefit from encoding syntactic relations, rather than linear sequences of lexical items.

We use the shortest path between entities in a dependency tree to capture the essential context.  Using the NewsReader pipeline, we identify the shortest path, and use the word embeddings for all tokens in the path as input to a neural network.  
Similar to previous work in relation extraction \cite{Xu2015}, we use two branches, where the left branch processes the path from the source entity to the least common ancestor (LCA), and the right branch processes the path from the target entity to the LCA.
However, our \textsc{tlink} extraction model uses only word embeddings as input, not POS tags, grammatical relations themselves, or WordNet hypernyms.

For example, for the sentence ``Their marriage ended before the war'', given an event pair ({\em marriage}, {\em war}), the left branch of the model will receive the sequence ({\em marriage, ended}), while the right branch will receive ({\em war, before, ended}).   
The LSTM layer processes the appropriate sequence of word embeddings in each branch.  This is followed by a separate max pooling layer for each branch, so for each LSTM unit, the maximum value over the time steps is used, not the final step value.
During the early stages of model design, we observed that this max pooling approach (also used in \newcite{Xu2015}) resulted in a slight improvement in performance.
Finally, the results from the max pooling layers of both branches are concatenated and fed to a hidden layer, followed by a softmax to yield a probability distribution over the classes.
The model architecture is shown in Figure \ref{fig:rnn_model} (left).
We also augment the training data by flipping every pair, 
i.e. if $(e_1, e_2)~\rightarrow$~\textsc{before}, $(e_2, e_1)~\rightarrow$~\textsc{after} 
is also included.

\subsection{Cross-Sentence Model}
\label{sect:cross}

\textsc{Tlink}s between the entities in consecutive sentences can often be identified without any external context or prior knowledge. For example, the order of events may be indicated by discourse connectives, or the events may follow natural order, potentially encoded in their word embeddings. 


To recover such relations, we use a model similar to the one used for intra-sentence relations, as described in Section\ref{sect:within}. 
Since there is no common root between entities in different sentences, we use the path between an entity and the sentence root to construct input data. A sentence root is often the main verb, or a conjunction. 



\subsection{Relations to DCT}
\label{dct}
The document creation time (DCT) naturally serves as the ``current time''. In this section, we discuss how to identify temporal relations between an event and DCT. The assumption here is that an event mention and its local context can often suffice for DCT \textsc{tlink}s. For example, English 
has inflected verbs for tense in finite clauses, and uses auxiliaries to express aspects. 



The model we use is again similar to the one in Section\ref{sect:cross}. 
Although one branch would suffice in this case, we use two branches in our implementation. One branch processes the path from a given entity to the sentence root, and the other branch processes the same path in reverse, from the root to the  entity. 


\subsection{Relations between \textsc{timex}es}
\label{sect:timex}
Time expressions explicitly signify a time point or an interval of time. Without the \textsc{timex} entities serving as ``hubs'', many events would be isolated from each other.  We use rule-based techniques to identify temporal relations between \textsc{timex} pairs that have been identified and normalized by HeidelTime.  
The relation between the DCT and other time expressions is just a special case of \textsc{timex}-to-\textsc{timex} \textsc{tlink} and is handled with rules as well. 

\begin{table}[h]
\begin{center}
\resizebox{\columnwidth}{!}{
\begin{tabular}{|l|l|l|}
\hline \bf \textsc{date} value & \bf Calculation & \bf Representation\\ \hline
2017-08-04 & START = $2017+7/12+3/365$ & (2017.591, 2017.591)\\
        & $=2017.591$ & \\ 
           & END = START & \\ \hline
2017-SU & START = $2017+5/12=2017.416$ & (2017.416, 2017.666)\\
(Summer 2017)  & END = $2017+8/12=2017.666$ & \\ \hline
\end{tabular}
}
\end{center}
\caption{\label{tab:data-values} Examples of \textsc{date} values and their tuple representations }
\end{table}

In the present implementation, we focus on the \textsc{date} class of \textsc{timex} tags, which is prevalent in the newswire text. The \textsc{time} class tags which contain more information are converted to \textsc{date}. Every \textsc{date} value is mapped to a tuple of real values (\textit{start}, \textit{end}). The ``value'' attribute of \textsc{timex} tags follows the ISO-8601 standard, so the mapping is straightforward. Table \ref{tab:data-values} provides some examples. 
We set the minimum time interval to be a day.
Practically, such a treatment suffices for our data.  After mapping \textsc{date} values to tuples of real numbers, we can define 5 relations between \textsc{timex} entities $T_1=(start_1, end_1)$ and $T_2=(start_2, end_2)$ as follows:

\begin{equation}\label{timex-relations}
\resizebox{0.8\hsize}{!}{$
    T_1\times T_2\rightarrow
    \begin{cases}
        \text{\textsc{before} }  & \text{if } end_1<start_2 \\
        \text{\textsc{after} }  & \text{if } start_1>end_2 \\
        \text{\textsc{includes} }  & \text{if } start_1<start_2 \\
                            & \text{ and } end_1>end_2 \\
        \text{\textsc{is\_included} }  & \text{if } start_1>start_2 \\
                            & \text{ and } end_1<end_2 \\
        \text{\textsc{simultaneous} }  & \text{if } start_1=start_2 \\
                            & \text{ and } end_1=end_2
    \end{cases} $
    }
\end{equation}

The \textsc{tlink}s from training data contain more types of relations than the five described in Equation  \ref{timex-relations}. 
However relations such as \textsc{ibefore} (``immediately before"), \textsc{iafter}(``immediately after") and \textsc{identity} are only used on event pairs, not \textsc{timex} pairs. The QA system also does not target questions on \textsc{timex} pairs. The purpose here is to use the \textsc{timex} relations to link the otherwise isolated events.


\section{Double-checking}
A major difficulty we have is that the \textsc{tlink}s for intra-sentence, cross-sentence, and DCT relations in the training data are not comprehensive. Often, the temporal relation between two entities is clear, but the training data provides no \textsc{tlink} annotation.
We downsampled the \textsc{no-link} class in training in order to address both the class imbalance and the fact that TimeML-style annotation is de-facto sparse, with only a fraction of positive instances annotated.

In addition to that, we introduce a technique to boost the recall of positive classes (not \textsc{no-link}) and to reduce the misclassification between the opposite classes.  Since entity pairs are always classified in both orders, if both orders produce a \textsc{tlink} relation, rather than \textsc{no-link}, we adopt the label with a higher probability score, as assigned by the softmax classifier. We call this technique ``double-checking". It serves to reduce the errors that are fundamentally harmful (e.g. \textsc{before} misclassified as \textsc{after}, and vice versa). 
We also allow a positive class to have the ``veto power'' against \textsc{no-link} class. 
For instance, if our model predicts $(e_1, e_2)\rightarrow \text{\textsc{after}}$ but \textsc{no-link} reversely, we adopt the former.


\begin{table}[h]
\begin{center}
\resizebox{\columnwidth}{!}{
\begin{tabular}{|@{} l|l|l|l|l @{}|}
\hline 
\bf \textsc{no-link} ratio & \bf Recall & \bf Recall & \bf \textsc{before} & \bf \textsc{after} \\
\bf               & \bf \textsc{before} & \bf \textsc{after} & \bf  as \textsc{after} & \bf as \textsc{before}\\
\hline
0.5  & 0.451 & 0.445 & 0.075 & 0.092\\
0.1  & 0.643 & 0.666 & 0.145 & 0.159\\
0.1 + double-check & 0.721 & 0.721 & 0.125 & 0.125\\
\hline
\end{tabular}
}
\end{center}
\vspace{-1em}
\caption{\label{witin-exp} 
Effects of downsampling and double-checking on intra-sentence results. 0.5 \textsc{no-link} ratio means that \textsc{no-link}s are downsampled to a half of the number of all positive instances combined. \textsc{Before} as \textsc{after} shows the fraction of \textsc{before} misclassified as \textsc{after}, and vice versa.}

\end{table}

Table \ref{witin-exp} shows the effects of double-checking and downsampling the \textsc{no-link} cases on the intra-sentence model.  
Double-checking technique not only further boosts recall, but also reduces the misclassification between the opposite classes.

\section{Pruning \textsc{Tlink}s}
\label{pruning}
The four \textsc{tlink} classification models in Section \ref{sec:tlink-classification} deal with different kinds of \textsc{tlink}s, so their output does not overlap. Nevertheless temporal relations are transitive in nature, so the deduced relations from given \textsc{tlink}s can be in conflict.

Most conflicts arise from two types of relations, namely \textsc{before}/\textsc{after} and \textsc{includes}/\textsc{is\_included}. Naturally, we can convert \textsc{tlink}s of opposite relations and put them all together. If we use a directed graph to represent the \textsc{before} relations between all entities, it should be acyclic.
Sun \shortcite{Sun:2014} proposed a strategy that ``prefers the edges that can be inferred by other edges in the graph and remove the ones that are least so". 
Another strategy is to use the results from separate classifiers or ``sieves'' to rank \textsc{tlink}s according to their confidence \cite{mani2007three,chambers2014dense}. High-ranking results overwrite low-ranking ones.


We follow the same idea of purging the weak \textsc{tlink}s. Given a directed graph, our approach is to remove the edges to break cycles, so that the sum of weights from the removed edges is minimal. 
This problem is actually an extension of the minimum feedback arc set problem and is NP-hard \cite{Karp:1972}.
We therefore adopt a heuristic-based approach, applied separately to the graphs induced by \textsc{before}/\textsc{after} and \textsc{includes}/\textsc{is\_included} relations.\footnote{We found that \textsc{ends} and \textsc{begins} \textsc{tlink}s are too infrequent to warrant a separate treatment.} 
The softmax layer provides a probability score for each relation class, which represents the strength of a link. \textsc{Tlink}s between \textsc{timex} pairs are generated by rules, so we assume them to be reliable and assign them a score of 1. 
Although \textsc{includes}/\textsc{is\_included} edges can generate conflicts in a \textsc{before}/\textsc{after} graph as well, we currently do not resolve such conflicts because they are relatively rare. We also do not use \textsc{simultaneous/identity} relations to merge nodes, because we found that it leads to very unstable results.

\begin{algorithm}[ht!] 
\footnotesize
 $X\gets$ EVENTS\; $V\gets$ TIMEXes\; $E\gets$ TIMEX pairs\;
 Initialize $G\gets <V, E>$\;
  \For{x$\in$ X}{
    $V'\gets V+\{x\}$\;
    $C \gets \{(x,v) \cup (v,x) | v\in V\}$ \;
    $E'\gets E\cup C$ \;
    $G'\gets <V',E'>$ \;
    \If{cycle$\_$exists(G')}{
    \For{$C_i \in \pi(C)$}{
      $score_i=0$\; 
     \While{$C_i\neq \phi \text{ \& } cycle\_exists(G\cup C_i)$}{
    $c\gets C_i.pop()$\;
    $score_i +=weight(c)$\;
    }
    }
       }
    $G\gets G\cup C_i$ s.t. $i=argmin(score_i)$\;
  }
  \caption{\small Algorithm to prune edges. $\pi(C)$ denotes some permutations of $C$, where $C$ is a list of weighted edges.} 
  \label{alg:pruning}
\end{algorithm}

For a given relation (e.g., \textsc{before}), we incrementally build a directed graph with all edges representing that relation.  We first initialize the graph with  \textsc{timex}-to-\textsc{timex} relations.
Event vertices are then added to this graph in a random order.
For each event, we add all edges associated with it. If this creates a cycle, the edges are removed one by one until there is no cycle, keeping track of the sum of the scores associated with removed edges.  We choose the order in which the edges are removed to minimize that value.\footnote{By removing an edge, we mean resetting the relation to \textsc{no-link}. Another possibility may be to set the relation associated with the edge to the one with the second highest probability score, however this may create additional cycles.}  The algorithm is shown above.
In practice, the vertices do not have a high degree for a given relation, so permuting the candidates $N\times(N-1)$ times (i.e., not fully), where $N$ is the number of candidates, 
produces only a negligible slowdown. We also make sure to try the greedy approach, dropping the edges with the smallest weights first.

%


\section{Model Settings}
\label{experiments}


In this section, we describe the model settings used in our experiments.
All models requiring word embeddings use 300-dimensional word2vec vectors trained on Google News corpus (3 billion running words).\footnote{\url{https://github.com/mmihaltz/word2vec-GoogleNews-vectors}} 
Our models are written in Keras on top of Theano. 



\paragraph{\textsc{Timex} and Event Annotation}


The LSTM layer of the event extraction model contains 128 LSTM units. The hidden layer on top of that has 30 neurons. The input layer corresponding to the 4 token features is connected with a hidden layer with 3 neurons. The combined hidden layer is then connected with a single-neuron output layer. We set a dropout rate 0.5 on input layer, and another drop out rate 0.5 on the hidden layer before output. 

As mentioned earlier, we do not attempt to tag event attributes.
Since the vast majority of tokens are outside of event mention boundaries, we set higher weights for the positive class. 
In order to answer questions about temporal relations, it is not particularly harmful to introduce spurious events, but missing an event makes it impossible to answer any question related to it. 
Therefore we intentionally boost the recall while sacrificing precision. 
Table \ref{entity-tags} shows the performance of our event extraction, as well as the performance of HeidelTime \textsc{timex} tagging. For events, partial overlap of mention boundaries is considered an error. 

\begin{table}[h!]
\begin{center}
\resizebox*{0.7\hsize}{!}{
\begin{tabular}{|l|l|l|l|}
\hline \bf Annotation & \bf Prec & \bf Rec & \bf F1\\ \hline
\textsc{timex} & 0.838 & 0.850 & 0.844\\
Event & 0.729 & 0.963 & 0.830\\
\hline
\end{tabular}
}
\end{center}
\vspace{-1em}
\caption{\label{entity-tags} \textsc{timex} and event evaluation on validation set. }
\end{table}

\paragraph{Intra-Sentence Model}
We identify 12 classes of temporal relations, plus a \textsc{no-link} class. For training, we downsampled \textsc{no-link} class to 10\% of the number of positive instances.
Our system does not attempt to resolve coreference.  For the purpose of identifying temporal relations, \textsc{simultaneous} and \textsc{identity} links capture the same relation of simultaneity, which allowed us to combine them.
%
%
%
The LSTM layer of the intra-sentence model contains 256 LSTM units on each branch. The hidden layer on top of that has 100 neurons. We set a dropout rate 0.6 on input layer, and another drop out rate 0.5 on the hidden layer before output. 
%

\paragraph{Cross-Sentence Model}

The training and evaluation procedures are very similar to what we did for intra-sentence models, and the hyperparameters for the neural networks are the same. Now the vast majority of entity pairs have no \textsc{tlink}s explicitly marked in training data. Unlike the intra-sentence scenario, however, a \textsc{no-link} label is truly adequate in most cases. 
We found that downsampling \textsc{no-link} instances to match the number of all positive instances (ratio=1) yields desirable results. Since positive instances are very sparse in both the training and validation data, the ratio should not be too low, so as not to risk overfitting.

\paragraph{DCT Model}
We use the same hyperparameters for the DCT model as for the intra-sentence and cross-sentence models. 
Again, the training files do not sufficiently annotate \textsc{tlink}s with DCT even if the relations are clear, so there are many false negatives. We downsample the \textsc{no-link} instances so that they are 4 times the number of positive instances.

\begin{table}[h]
\begin{center}
\resizebox*{\hsize}{!}{
\begin{tabular}{|l|c|c|c|c|c|c|c|}
\hline 
\bf system	&\bf coverage &\bf prec &\bf rec &\bf f1 \\ \hline
human-fold1-original		&0.43	&0.91	&0.38	&0.54 \\
human-fold1-timlinks		&0.52	&0.93	&0.47	&\bf 0.62 \\
TIPSem-fold1-original		&0.35	&0.57	&0.22	&0.32\\
TIPSem-fold1-timex		&0.53	&0.69	&0.38	&0.50\\ \hline
orig. validation data		&0.37	&\bf 0.93	&0.34	&0.50\\
orig. tags TEA tlinks		&0.81	&0.58	&0.47	&0.52\\
TEA-initial	&0.78	&0.60	&0.47	&0.52\\
TEA-double-check		&\bf 0.89	&0.60	&\bf 0.53	&0.56\\
TEA-prune		&0.82	&0.58	&0.48	&0.53\\ \hline
TEA-flat		&0.81	&0.44	& 0.35	& 0.39\\
TEA-Dense   &0.68	&0.70	& 0.48	& 0.57\\ \hline
TEA-final		&0.84	&0.64	&\bf 0.53	&\bf 0.58\\
\hline
\end{tabular}
}
\end{center}
\vspace{-1em}
\caption{\label{results-val} QA results on validation data. There are $\mathbf{79}$ questions in total. The 4 systems on the top of the table are provided with the toolkit. The systems starting with ``human-" are annotated by human experts. TEA-final 
utilizes both double-check and pruning. TEA-flat 
uses the
flat context. TEA-Dense is trained on TimeBank-Dense.}
\end{table}

\section{Experiments}
\label{results}

In this section, we first describe the model selection experiments on QA-TempEval validation data, selectively highlighting results of interest.  We then present the results obtained with the optimized model on the QA-TempEval task and on TimeBank-Dense.

\subsection{Model Selection Experiments}
%
%
%
%
As mentioned before, ``gold'' \textsc{tlink}s are sparse, so we cannot merely rely on the F1 scores on validation data to do model selection. Instead, we used the QA toolkit. 
The toolkit contains 79 yes-no questions about temporal relations between entities in the validation data. Originally, only 6 questions have ``no" as the correct answer, and 1 question is listed as ``unknown".  After investigating the questions and answers, however, we found some errors and typos\footnote{Question 24 from \textsf{XIE19980821.0077.tml} should be answered with ``yes", but the answer key contains a typo ``is". Question 34 from \textsf{APW19980219.0476.tml} has \textsc{before} that should be replaced with \textsc{after}. Question 29 from \textsf{XIE19980821.0077.tml} has ``unknown" in the answer key, but after reading the article, we believe the correct answer is ``no".}. After fixing the errors, there are 7 no-questions and 72 yes-questions in total. All evaluations are performed on the fixed data.

The evaluation tool draws answers from the annotations only. If an entity (event or \textsc{timex}) involved in a question is not annotated, or the \textsc{tlink} cannot be found, the question will then be counted as not answered. There is no way for participants to give an answer directly, other than delivering the annotations. The program generates Timegraphs to infer relations from the annotated \textsc{tlink}s. As a result, relations without explicit \textsc{tlink} labels can still be used if they can be inferred from the annotations. The QA toolkit uses the following evaluation measures: 
\begin{description}
\item \small coverage = \normalsize $\frac{\# \text{answered}}{\# \text{questions}}$, \small precision = \normalsize $\frac{\# \text{correct}}{\# \text{answered}}$
\item \small recall = \normalsize $\frac{\# \text{correct}}{\# \text{questions}}$, \small f1 = \normalsize $\frac{2\times \text{precision}\times \text{recall}}{\text{precision}+\text{recall}}$
\end{description}

Table \ref{results-val} shows the results produced by different models on the validation data. The results of the four systems above the first horizontal line are provided by the task organizer. Among them, the top two use annotations provided by human experts. As we can see, the precision is very high, both above 0.90. Our models cannot reach that precision. In spite of the lower precision, automated systems can have much higher coverages i.e. answer a lot more questions. 

As a starting point, we evaluated the validation files in their original form, and the results are shown as ``orig. validation data" of Table \ref{results-val}. The precision was good, but with very low coverage. This supports our claim that the \textsc{tlink}s provided by the training/validation files are not complete. We also tried using the event and \textsc{timex} tags from the validation data, but performing \textsc{tlink} classification with our system. As shown with ``orig. tags TEA tlinks" in the table, now the coverage rises to 64 (or 0.81), and the overall F1 score reaches 0.52. The TEA-initial system uses our own annotators. The performance is similar, with a slight improvement in precision. This result shows our event and \textsc{timex} tags work well, and are not inferior to the ones provided by the training data. 

The double-checking technique boosts the coverage a lot, probably because we allow positive results to veto \textsc{no-link}s. Combining double-checking with the pruning technique yields the best results, with F1 score 0.58, answering 42 out of 79 questions correctly.

In order to validate the choice of the dependency path-based context, we also experimented with a conventional flat context window, using the same hyperparameters. Every entity is represented by a 11-word window, with the entity mention in the middle. If two entities are near each other, their windows are cut short before reaching the other entity. Using the flat context instead of dependency paths yields a much weaker performance. This confirms our hypothesis that syntactic dependencies represent temporal relations better than word windows. However, it should be noted that we did not separately optimize the models for the flat context setting. 
The large performance drop we saw from switching to flat context did not warrant performing a separate parameter search.

We also wanted to check whether a comprehensive annotation of \textsc{tlink}s in the training data can improve model performance on the QA task.
We therefore trained our model on TimeBank-Dense data and evaluated it with QA (see the TEA-Dense line in Table \ref{results-val}).
Interestingly, the performance is nearly as good as our top model, although TimeBank-Dense only uses five major classes of relations. For one thing, it shows that our system may perform equally after being trained on sparsely labeled data and on densely labeled data, judged from the QA evaluation tool. If this is true, excessively annotated data may not be necessary in some tasks.

\begin{table}[h!]
\begin{center}
\resizebox*{0.9\hsize}{!}{
\begin{tabular}{|c|c|c|c|c|c|c|c|}
 \hline 
	&\bf doc &\bf words &\bf quest &\bf yes &\bf no &\bf dist- &\bf dist+ \\ \hline
 news	&10	&6920	&99	&93	&6	&40	&59 \\
 wiki	&10	&14842	&130	&117	&13	&58	&72 \\
 blogs	&8	&2053	&65	&65	&0	&30	&35 \\ \hline
 total	&28	&23815	&294	&275	&19	&128	&166 \\
\hline
\end{tabular}
}
\end{center}
\vspace{-1em}
\caption{Test data statistics. Adapted from Table 1 in  \newcite{Llorens2015:TempEval}.}
\label{test-data}
\end{table}


\subsection{QA-TempEval Experiments}

We use the QA toolkit provided by the QA-TempEval organizers to 
evaluate our system on the test data. 
The documents in test data are not annotated at all, so the event tags, \textsc{timex} tags, and \textsc{tlink}s are all created by our system. 

Table \ref{test-data} shows the the statistics of test data. 
As we can see, the vast majority of the questions in the test set should be answered with $yes$. Generally speaking, it is much more difficult to validate a specific relation (answer $yes$) than to reject it (answer $no$) when we have as many as 12 types of relations in addition to the vague \textsc{no-link} class. \textbf{dist-} means questions involving entities that are in the same sentence or in consecutive sentences. \textbf{dist+} means the entities are farther away.

The QA-TempEval task organizers used two evaluation methods. The first method is exactly the same as the one we used on validation data. The second method used a so-called Time Expression Reasoner (TREFL) to add relations between \textsc{timex}es, and evaluated the augmented results. The goal of such an extra run is to ``analyze how a general time expression reasoner could improve results". Our  model already includes a component to handle \textsc{timex} relations, so we will compare our results with other systems' in both methods.

\begin{table}[h!]
\begin{center}
\resizebox*{\hsize}{!}{
\begin{tabular}{|l|c|c|c||c|c|}

\multicolumn{6}{c}{News Genre (99 questions)} \\
\hline 
\bf system	&\bf prec &\bf rec &\bf f1 &\bf \% answd &\bf \# correct   \\ \hline
hlt-fbk-ev1-trel1 &0.59	&0.17	&0.27   &29 &17 \\
hlt-fbk-ev1-trel2 &0.43	&0.23	&0.30   &55 &23 \\
hlt-fbk-ev2-trel1 &0.56	&0.20	&0.30   &36 &20 \\
hlt-fbk-ev2-trel2 &0.43	&0.29	&0.35   &69 &29 \\
ClearTK &0.60	&0.06	&0.11   &10 &6 \\
CAEVO &0.59	&0.17	&0.27   &29 &17 \\
TIPSemB &0.50	&0.16	&0.24   &32 &16 \\
TIPSem & 0.52	&0.11	&0.18   &21 &11 \\ \hline
TEA &\bf 0.61	&\bf 0.44	&\bf 0.51   &\bf 73 &\bf 44 \\
\hline
\end{tabular}
}
\newline
\vspace*{0.15cm}
\resizebox*{\hsize}{!}{
\begin{tabular}{|l|c|c|c||c|c|}

\multicolumn{6}{c}{Wikipedia Genre (130 questions)} \\
\hline 
\bf system	&\bf prec &\bf rec &\bf f1 &\bf \% answd &\bf \# correct   \\ \hline
hlt-fbk-ev1-trel1 &0.55	&0.16	&0.25   &29 &21 \\
hlt-fbk-ev1-trel2 &0.52	&0.22	&0.35   &50 &34 \\
hlt-fbk-ev2-trel1 &0.58	&0.17	&0.26   &29 &22 \\
hlt-fbk-ev2-trel2 &0.62	&0.36	&0.46   &58 &47 \\
ClearTK &0.60	&0.05	&0.09   &8 &6 \\
CAEVO &0.59	&0.17	&0.26   &28 &22 \\
TIPSemB &0.52	&0.13	&0.21   &25 &17 \\
TIPSem &\bf 0.74	&0.19	&0.30   &26 &25 \\ \hline
TEA &0.62	&\bf 0.44	&\bf 0.51  &\bf 71 &\bf 57 \\
\hline
\end{tabular}
}
\newline
\vspace*{0.15 cm}
\resizebox*{\hsize}{!}{
\begin{tabular}{|l|c|c|c||c|c|}

\multicolumn{6}{c}{Blog Genre (65 questions)} \\
\hline 
\bf system	&\bf prec &\bf rec &\bf f1 &\bf \% answd &\bf \# correct   \\ \hline
hlt-fbk-ev1-trel1 &\bf 0.57	&0.18	&\bf 0.28   &32 &12 \\
hlt-fbk-ev1-trel2 &0.43	&0.18	&0.26   &43 &12 \\
hlt-fbk-ev2-trel1 &0.47	&0.14	&0.21   &29 &9 \\
hlt-fbk-ev2-trel2 &0.34	&\bf 0.20	&0.25   &\bf 58 &\bf 13 \\
ClearTK &0.56	&0.08	&0.14   &14 &5 \\
CAEVO &0.48	    &0.18	&0.27   &38 &12 \\
TIPSemB &0.31	&0.08	&0.12   &25 &5 \\
TIPSem & 0.45	&0.14	&0.21   &31 &9 \\ \hline
TEA &0.43	&\bf 0.20	&0.27  & 46 &\bf 13 \\
\hline
\end{tabular}
}
\end{center}
\vspace{-1em}
\caption{QA evaluation on test data without TREFL}
\label{results-test}
\end{table}

The results are shown in Table \ref{results-test}. We give the results for the hlt-fbk systems that were submitted by the top team.
Among them, hlt-fbk-ev2-trel2 was the overall winner of TempEval task in 2015. ClearTK, CAEVO, TIPSEMB and TIPSem were some off-the-shelf systems provided by the task organizers for reference. These systems were not optimized for the task \cite{Llorens2015:TempEval}. 

For news and Wikipedia genres, our system outperforms all other systems by a large margin. 
For blogs genre, however, the advantage of our system is unclear. 
Recall that our training set contains news articles only. While the trained model works well on Wikipedia dataset too, blog dataset is fundamentally different in the following ways: (1) each blog article is very short, (2) the style of writing in blogs is much more informal, with non-standard spelling and punctuation, and (3) blogs are written in first person, and the content is usually personal stories and feelings. 

Interestingly, the comparison between different hlt-fbk submissions suggests that resolving event coreference (implemented by hlt-fbk-ev2-trel2) substantially improves system performance for the news and Wikipedia genres.  However, although our system does not attempt to handle event coreference explicitly, it easily outperforms the hlt-fbk-ev2-trel2 system in the genres where coreference seems to matter the most.  
%
%
\paragraph*{Evaluation with TREFL}
The extra evaluation with TREFL has a post-processing step that adds \textsc{tlink}s between \textsc{timex} entities. Our model already employs such a strategy, so this post-processing does not help. In fact, it drags down the scores a little. 
\begin{table}
\resizebox*{\hsize}{!}{
\begin{tabular}{|l|c|c|c||c|c|}
\multicolumn{6}{c}{All Genres (294 questions)} \\
\hline 
\bf system	&\bf prec &\bf rec &\bf f1 &\bf \% awd &\bf \# corr   \\ \hline
hlt-fbk-ev2-trel2   &0.49	&0.30	&0.37	&62	&89 \\
hlt-fbk-ev2-trel2-TREFL &0.51	&0.34	&0.40   &\bf 67 &99 \\\hline 
TEA &\bf 0.59	&\bf 0.39	&\bf 0.47   &66 &\bf 114 \\
TEA-TREFL &0.58	&0.38	&0.46   &66 &111 \\
\hline
\end{tabular}
}
\vspace{-1em}
\caption{Test results over all genres. 
}
\label{results-all}
\end{table}
Table \ref{results-all} summarizes the results over all genres before and after applying TREFL. 
For comparison, we include the top 2015 system, hlt-fbk-ev2-trel2. As we can see, TEA generally shows substantially 
higher scores.
%

\subsection{TimeBank-Dense Experiments}
We trained and evaluated the same system on TimeBank-Dense to see how it performs on a similar task with a different set of labels and another method of evaluation. In this experiment, we used the event and \textsc{timex} tags from test data, as \newcite{CATENA}.

Since all the \textsc{no-link} (vague) relations are labeled, downsampling was not necessary. We did use double-checking in the final conflict resolution, but without giving positive cases the veto power over \textsc{no-link}. Because \textsc{no-link} relations dominate, especially for cross-sentence pairs, we set class weights to be inversely proportional to the class frequencies during training. 
We also reduced input batch size to counteract class imbalance.

We ran two sets of experiments. One used the uniform configurations for all the neural network models, similar to our experiments with QA-TempEval. The other tuned the hyperparameters for each component model (number of neurons, dropout rates, and early stop) separately. 

The results from TimeBank-Dense are shown in Talble~\ref{results-dense}. 
Even though TimeBank-Dense has 
a very different methodology for both annotation and evaluation,
our ``out-of-the-box'' model which uses uniform configurations across different components obtains F1 0.505, compared to the best F1 of 0.511 in previous work.  Our best result of 0.519 is obtained by tuning hyperparameters on intra-sentence, cross-sentence, and DCT models independently.




For the QA-TempEval task, we intentionally tagged a lot of events, and let the pruning algorithm resolve potential conflicts. In the TimeBank-Dense experiment, however, we only used the provided event tags, which are sparser than what we have in QA-TempEval. The system may have lost some leverage that way.

\begin{table}
\resizebox*{\hsize}{!}{
\begin{tabular}{|l||c|c|c|c|c|c|}
\hline 
\bf system	&\bf ClearTK &\bf NavyT &\bf CAEVO &\bf CATENA & \multicolumn{2}{|c|}{\bf TEA-Dense}   \\ \hline
\bf  &  &  &  & &\bf uniform & \bf tuned   \\ 
F1   &0.447	&0.453	&0.507	&0.511	&0.505 & 0.519 \\
\hline
\end{tabular}
}
\vspace{-1em}
\caption{TEA results on TimeBank-Dense.
ClearTK, NavyT, and CAEVO are systems from \newcite{chambers2014dense}.  CATENA is from \newcite{CATENA} }
\label{results-dense}
\end{table}

\section{Conclusion}
\label{conclusion}


We have proposed a new method for extraction of temporal relations which takes a relatively simple LSTM-based architecture, using shortest dependency paths as input, and re-deploys it in a set of subtasks needed for extraction of temporal relations from text. We also introduce two techniques that leverage confidence scores produced by different system components to substantially improve the results of \textsc{tlink} classification:
(1) a ``double-checking'' technique which reverses pairs in classification, thus boosting the recall of positives and reducing misclassifications among opposite classes and (2) an efficient pruning algorithm to resolve \textsc{tlink} conflicts.
%
In a QA-based evaluation, our proposed method 
outperforms state-of-the-art methods by a large margin.
We also obtain state-of-the art results in an intrinsic evaluation on a very different TimeBank-Dense dataset, proving generalizability of the proposed model.




\section*{Acknowledgments}
This project is funded in part by an NSF CAREER Award to Anna Rumshisky (IIS-1652742). We would like to thank Connor Cooper and Kevin Wacome for their contributions to the early stages of this work.

\bibliography{emnlp2017}
\bibliographystyle{emnlp_natbib}

\end{document}